\documentclass[aps,pre,showpacs,foatfix,nobibnotes,footinbib,
               superscriptaddress,preprint]{revtex4}

\usepackage{amsmath,amssymb,amsfonts}
\usepackage{graphicx}
\usepackage[active]{srcltx} 

\newcommand{\dif}{\partial}

\newcommand{\ev}[1]{\langle {#1} \rangle}

\DeclareMathOperator{\const}{const}

\newcommand{\al}{\alpha}     \newcommand{\be}{\beta}
\newcommand{\ga}{\gamma}     \newcommand{\de}{\delta}
\newcommand{\ep}{\epsilon}   \newcommand{\ve}{\varepsilon}

\newcommand{\la}{\lambda}    \newcommand{\om}{\omega}
   \newcommand{\si}{\sigma}
  \newcommand{\te}{\theta}
  \newcommand{\ze}{\zeta}

\newcommand{\De}{\Delta}

\begin{document}

\title{Surface diffusion coefficient near first-order phase transitions at low temperatures}

\author{Igor Medved'}
\email{imedved@ukf.sk}
\author{Anton Trn\'{\i}k}
\affiliation{Department of Physics, Constantine the Philosopher
University, 94974 Nitra, Slovakia}
\affiliation{Department of Materials Engineering and Chemistry, Czech Technical University, 16629 Prague, Czech Republic}

\begin{abstract}
We analyze the collective surface diffusion coefficient, $D_c$, near a first-order phase transition at which two phases coexist and the surface coverage, $\te$, drops from one single-phase value, $\te_+$, to the other one, $\te_-$. Contrary to other studies, we consider the temperatures that are sufficiently sub-critical. Using the local equilibrium approximation, we obtain, both numerically and analytically, the dependence of $D_c$ on the coverage and system size, $N$, near such a transition. In the two-phase regime, when $\te$ ranges between $\te_-$ and $\te_+$, the diffusion coefficient behaves as a sum of two hyperbolas, $D_c \approx A/N|\te - \te_-| + B/N|\te - \te_+|$. The steep hyperbolic increase in $D_c$ near $\te_\pm$ rapidly slows down when the system gets from the two-phase regime to either of the single-phase regimes (when $\te$ gets below $\te_-$ or above $\te_+$), where it approaches a finite value. The crossover behavior of $D_c$ between the two-phase and single-phase regimes is described by a rather complex formula involving the Lambert function. We consider a lattice-gas model on a triangular lattice to illustrate these general results, applying them to four specific examples of transitions exhibited by the model.
\end{abstract}


\maketitle


\section{Introduction}

The collective (or chemical) surface diffusion coefficient, $D_c$, is defined via the Fick's first law and represents a relevant transport coefficient for surface diffusion. Theoretical studies of $D_c$ and of the influence of lateral interparticle interactions on $D_c$ have often used lattice-gas models to simulate surface diffusion. In the models the migration of adparticles is given by the potential relief of the substrate surface: most of the time the adparticles stay at the positions (sites) where the relief attains its minima, but from time to time they perform random jumps to the adjacent vacant sites. Assuming the jumps to be instant, the states of the system of adparticles are represented by the occupation numbers (one for each site), as in a lattice gas. Although this description is rather oversimplifying, it should possess the key aspects of the diffusion and, moreover, it can be treated by a number of statistical mechanical methods, such as the mean-field, real-space renormalization group, and computer simulation techniques \cite{Ala02,Nau05}.

In order to determine $D_c$ in general, one should solve a system of balance equations for a large number of adparticles that may strongly interact with each other as well as with the substrate surface. An analytic treatment of such a formidable kinetic problem often resorts to some kind of approximation. In particular, assuming that the adparticle surface coverage varies only very slowly with time and space (the local equilibrium limit), purely thermodynamic quantities are sufficient to obtain $D_c$, i.e., the problem reduces to the evaluation of the finite-size specific free energy, $f$, of the system \cite{Tar80,Re81a,Re81,Zh91}. Namely, assume that the jumps of adparticles are mutually uncorrelated and restricted to nearest neighbors. In addition, assume that an activated adparticle at a saddle point of the potential barrier interacts only with the nearest-neighbor adparticles. Then the original problem can be reduced to a diffusion equation, with the corresponding diffusion coefficient given as \cite{Tar80,Ta01,Ta03}
\begin{equation} \label{eq: D_c}
  D_c \approx D_c^0 \, e^{\be\mu} \frac P{\chi/\be}.
\end{equation}
Here $D_c^0$ is the diffusion coefficient of non-interacting particles, $\be = 1/k_B T$ is the inverse temperature, $\mu$ is the chemical potential, $\chi$ is the isothermal susceptibility, and $P$ is a correlation factor. Both $\chi$ and $P$ can be expressed as derivatives of $f$ [see Eqs.~\eqref{eq: P gen} and \eqref{eq: quant gen} below]. For lattice gases the local equilibrium approximation that leads to an expression for $D_c$ given only via thermodynamic quantities turns out to be fairly plausible: the results obtained from it by the analytical methods have been in quite good agreement with the numerical results obtained by kinetic simulations \cite{Ta07a}.

The correlation factor $P$ is associated with the interactions of an activated adparticle with other particles. It is given as a sum of the probabilities that certain clusters of adjacent sites are vacant \cite{Ta01,Ta07a,Ta09}. The clusters contain a lattice bond representing the two sites between which a particle jump is performed, plus the neighboring sites with which an activated adparticle is supposed to interact. Usually, only the sites nearest to the saddle point are considered. Then the clusters are quite small; for example, for a triangular lattice these are only bonds and elementary triangles and parallelograms \cite{Ta01}. Clearly, the probabilities that clusters are vacant may be expressed via derivatives of $f$ with respect to suitable interparticle interaction parameters, $p_i$. Therefore, quite generally, $P$ has the form given as
\begin{equation} \label{eq: P gen}
  P = a_0 + \sum_i a_i \frac{\dif f}{\dif p_i},
\end{equation}
where the constants $a_i$ may depend on the interaction parameters of an activated adparticle.

One of the intriguing problems that has attracted particular attention is the presence of phase transitions and their effects on surface diffusion. Since lattice gases can be used to model such transitions, they have provided a convenient framework also in this regard \cite{UG91,UG95,Zh95,Ta01,Ta03,Za04,Chv06,Za07,Mas07,Chv08,Ta08,Za10}. However, below critical temperatures ordered phases may arise due to lateral interactions, and sophisticated arguments should be applied to analyze surface diffusion \cite{Ta01}. In fact, very low temperatures have not been considered in the previous studies.

In this paper we wish to fill in this gap and study the diffusion coefficient $D_c$ at all sufficiently low temperatures, concentrating on its dependence on the surface coverage, chemical potential, and system's size. Our analysis is based on two key points. First, we assume that $D_c$ can be approximated by the expression~\eqref{eq: D_c}, which is appropriate only in the local equilibrium limit and under the above-mentioned restrictions on the adparticle jumps. Then $D_c$ can be obtained just from the finite-size specific free energy $f = - (1/\be N) \ln Z $ of the system, where $N$ is the number of adparticles in the system and $Z$ is its finite-size partition function. Second, we assume that a first-order phase transition between two low-temperature phases, $p_+$ and $p_-$, takes place in the system at a transition point $\mu = \mu_t$. Consequently, we will be able to employ the formula \cite{MT12,BoKo90}
\begin{equation} \label{eq: Z}
  Z (\mu) =  [ \nu_- e^{-\be f_- (\mu) N}
  + \nu_+ e^{-\be f_+ (\mu) N} ] (1 + r)
\end{equation}
that is applicable near the transition point $\mu_t$ for a large class of lattice-gas models with periodic boundary conditions, such as the models with a finite range $m$-potential and a finite number of ground states. Here $\nu_\pm$ and $f_\pm$ is the degeneracy and single-phase specific free energy of phase $p_\pm$, respectively, and the error term $r = O[\exp(- \const \be \sqrt N)]$. [The symbol $O(x)$ represents a term that can be bounded by $\const x$.] Combining Eqs.~\eqref{eq: D_c} to \eqref{eq: Z}, we will obtain analytic formulas for the dependence of $D_c$ on the chemical potential and coverage near the transition.

We find it useful to illustrate the general results with a specific lattice-gas model. Therefore, we will consider a model on a regular triangular lattice in which phase transitions between ordered phases can occur. Moreover, instead of the general form~\eqref{eq: P gen} of the correlation factor $P$, we will work with the widely used form in which $P$ is identified with the probability that a lattice bond is vacant. This corresponds to the simplest case when an activated adparticle does not interact with any neighbors. Then, for the triangular lattice, one has \cite{Ta01,Ta03}
\begin{equation} \label{eq: P}
  P = 1 - 2\te + \frac13 \xi,
\end{equation}
where $\te$ and $\xi$ is the statistical average number (per site) of occupied sites and bonds, respectively. (Note that $\te$ is the surface coverage.) We will eventually show that our results can be, after additional analysis, extended to the general form~\eqref{eq: P gen} of $P$.

The paper has the following structure. In Sec.~\ref{sec: model} the illustrative model of surface diffusion on the triangular lattice is introduced and its low temperature phases and free energy $f$ are described. The coverage dependence of $D_c$ is analyzed in Sec.~\ref{sec: Dc}. The derivation of general finite-size formulas for this dependence is presented, and  the general results are applied to the considered model. Concluding remarks, including the extension of the obtained results to the general form of $P$, are given in a final section.

\section{The model} \label{sec: model}

The model assumes that particles can be adsorbed on a solid surface only at sites forming a regular triangular lattice. The system contains a rectangular array with a large but finite number, $N$, of adsorption sites. Periodic boundary conditions are applied so that the array forms a finite torus. Setting the mesh size equal to $1$, the elementary lattice vectors are taken as $(1,0)$ and $(1/2, \sqrt 3/2)$. The torus cell is specified by the vectors $(3n,0)$ and $(0,2\sqrt 3 n)$ with $n = 1, 2, \dots$; thus, $N = 3n \times 4n$.

Each lattice site is either vacant or occupied by a particle. The interaction between two particles is limited to nearest-neighbor pairs (bonds) with an interaction energy that depends on the surrounding particles in the simplest possible way---only on the presence of particles at the sites closest to the bond. For the triangular lattice there are two such sites. The bond together with either of the sites forms an elementary triangle. Hence, the varying interaction energy is equivalent to having two constant interaction energies: one, $\ve_b$, for occupied bonds and one, $\ve_t$, for occupied elementary triangles. The corresponding Hamiltonian is given as \cite{Ta01,Ta03}
\begin{equation} \label{eq: H}
  H = \ve_b N_b + \ve_t N_t - \mu N_s,
\end{equation}
where $N_b$, $N_t$, and $N_s$ is the number of occupied bonds, elementary triangles, and sites, respectively. This model was already used to study surface diffusion at high temperatures in the special cases when $\ve_b = 0$ (for $T$ above $0.21 |\ve_t|/k_B$) and $\ve_t = 0$ (with $T$ above $0.1 |\ve_b|/k_B$) \cite{Ta03}. Here we consider the general case when both the bond and triangle interactions $\ve_b$ and $\ve_t$ are present, and temperatures are supposed to be sufficiently low.

\begin{figure}
  \centering
  \includegraphics{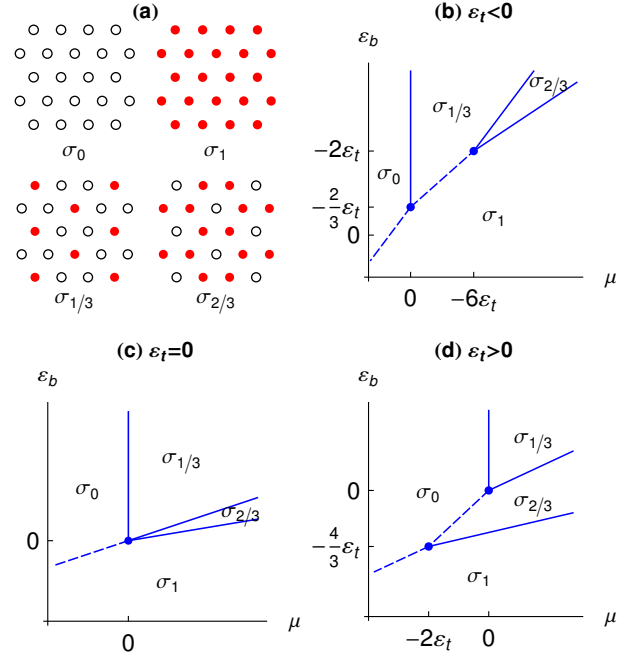}\\
\caption{(a) The ground states of the model. Circles (disks) represent vacant (occupied) sites. (b)--(d) The ground-state diagram in dependence on the sign (attractivity or repulsivity) of the triangle interaction $\ve_t$. On the boundaries between two ground-state regions either the two ground states coexist (the dashed lines) or there are infinitely many ground states (the solid lines, including their end-points depicted by disks).}
  \label{fig: GS}
\end{figure}

As we proved in \cite{MT12}, model~\eqref{eq: H} has four ground states [see Fig.~\ref{fig: GS}(a)]: a fully vacant state, $\si_0$, a fully occupied state, $\si_1$, and two threefold degenerate states, $\si_{1/3}$ and $\si_{2/3}$. The coverage of the two latter states is only partial, namely, $1/3$ and $2/3$, respectively. The ground-state diagram is shown in Fig.~\ref{fig: GS}(b)--(d) and can be easily constructed by comparing the four ground-state energies $e_0 = 0$,  $e_1 = 3\ve_b + 2\ve_t - \mu$, $e_{1/3} = -\mu/3$, and $e_{2/3} = \ve_b - 2\mu/3$. On the lines separating the regions of ground states $\si_0$ and $\si_1$, $\si_0$ and $\si_{2/3}$, and $\si_1$ and $\si_{1/3}$ (the dashed lines in Fig.~\ref{fig: GS}), only these two ground states coexist. However, on the remaining lines (the solid lines in Fig.~\ref{fig: GS}) as well as at the points where three or all four ground-state regions meet, there is an infinite number of ground states, yielding in fact a residual entropy.

Each ground state $\si_\al$, $\al = 0, 1/3, 2/3, 1$, gives rise to a unique low-temperature phase, $p_\al$, whose typical configuration looks as a `sea' of the ground state $\si_\al$ in which isolated `islands' of non-ground-state configurations are scattered, thus resembling the structure of $\si_\al$. So, phase $p_0$ ($p_1$) is fully vacant (fully occupied), while phase $p_{1/3}$ ($p_{2/3}$) has the occupancy of $1/3$ ($2/3$). The existence of these low-temperature phases can be concluded only if the number of ground states is finite, i.e., only within each ground-state region and on the lines between these regions where only two ground states coexist (the dashed lines in Fig.~\ref{fig: GS}). Otherwise, no conclusions concerning low-temperature phases were drawn in \cite{MT12}. Consequently, a first-order phase transition can take place between phases $p_0$ and $p_1$, $p_0$ and $p_{2/3}$, and $p_{1/3}$ and $p_1$, whereas transitions between other phases need not be of first-order.

Let us consider a pair of phases, $p_-$ and $p_+$ , between which a first-order phase transition occurs (specific pairs of phases will be considered later in Sec.~\ref{sec: Dc}). The associated ground states are denoted as $\si_-$ and $\si_+$, respectively. At low temperatures and near the transition point $\mu_t$ (for $|\mu - \mu_t| \leq \const / \be \sqrt N$), the finite-size partition function, $Z$, can be expressed as a sum of two single-phase partition functions, leading to Eq.~\eqref{eq: Z} for $Z$. This in turn yields the finite-size specific free energy as \cite{MT12}
\begin{equation} \label{eq: f}
  f (\mu) =  - \frac1{\be N}
  \ln [ \nu_- e^{-\be f_- (\mu) N} + \nu_+ e^{-\be f_+ (\mu) N} ]
  + r.
\end{equation}
The degeneracy $\nu_\pm$ for model~\eqref{eq: H} is equal to $1$ if $p_\pm$ is phase $p_0$ or $p_1$ and to $3$ if $p_\pm$ is phase $p_{1/3}$ or $p_{2/3}$.

The single-phase specific free energies $f_\pm$ are essentially equal to the ground-state specific energies, $e_\pm$, because the contributions from the thermal perturbations of the ground states $\si_\pm$ are suppressed exponentially in $\be$ (the Peierls condition). Namely, taking into account only one-site perturbations (which represent the leading corrections), one has \cite{MT12}
\begin{equation} \label{eq: f_al}
  f_\al \approx e_\al - \frac1\be \,
  \ln [(1 + e^{ - \be \, \De H_\al^\circ })^\al
       (1 + e^{ - \be \, \De H_\al^\bullet })^{1-\al}],
\end{equation}
where
\begin{equation}
\begin{aligned}
  \De H_0^\bullet &= - \De H_{1/3}^\circ = - \mu,
\\
  \De H_{1/3}^\bullet &= - \De H_{2/3}^\circ = 3\ve_b - \mu,
\\
  \De H_{2/3}^\bullet &= - \De H_1^\circ = 6\ve_b + 6\ve_t -\mu
\end{aligned}
\end{equation}
are the energy excesses of one-site perturbations of $\si_\al$ over $\si_\al$  (the superscript `$\circ$' corresponds to removing one particle from $\si_\al$ and `$\bullet$' to adding one particle to $\si_\al$). Since a particle can be only added to $\si_0$ (removed from $\si_1$), in Eq.~\eqref{eq: f_al} for $\al = 0$ ($\al = 1$) the first (second) term in $\ln$ is set equal to $1$.

The finite-size specific free energy $f$ of the model can be readily evaluated from Eqs.~\eqref{eq: f} and \eqref{eq: f_al}. The value of the transition point $\mu_t$ can be also obtained---it is the solution of the equation $f_- (\mu_t) = f_+ (\mu_t)$ \cite{BoKo90}.

\section{The coverage dependence of $D_c$} \label{sec: Dc}

Relations~\eqref{eq: D_c} and \eqref{eq: P} allows us to calculate an approximate value of the diffusion coefficient $D_c$ in the local equilibrium limit from the free energy $f$. Indeed, it suffices to find the derivatives \begin{equation} \label{eq: quant gen}
  \te = - \frac{\dif f}{\dif \mu},
  \quad
  \chi = - \frac{\dif^2 f}{\dif \mu^2},
  \quad
  \xi = \frac{\ev{N_b}}N = \frac{\dif f}{\dif \ve_b}.
\end{equation}
Hence, in combination with Eq.~\eqref{eq: f} that holds analogously also for derivatives of $f$ \cite{BoKo90}, the dependence of $D_c$ on $\mu$ immediately follows. Consequently, the coverage dependence of $D_c$ follows upon obtaining the inverse to $\te (\mu)$ and substituting it into $D_c (\mu)$, which can be easily carried out numerically.

However, we can also derive explicit finite-size formulas for $D_c (\te)$, starting from the $\mu$ dependences of $\te$, $\xi$, and $\chi$ as yielded by Eqs.~\eqref{eq: f} and \eqref{eq: quant gen}. Without loss of generality, we will assume that phase $p_-$ ($p_+$) is stable for $\mu$ below (above) $\mu_t$. Then the coverage jump at the transition is $\De\te = \te_+ - \te_- > 0$, where $\te_\pm = (- \dif f_\pm / \dif\mu)_{\mu_t}$ are the single-phase coverages at the transition.

\emph{Notation.} As a rule, we will use $\De q$ to denote the difference, $q_+ - q_-$, of single-phase quantities $q_+$ and $q_-$.

It turns out that three regimes in the behavior of $D_c (\te)$ may be distinguished according to the relative importance of phases~$p_-$ and $p_+$ as given by the weights
\begin{equation} \label{eq: la}
  \la_\pm (\mu) =
  \frac{ \nu_\pm e^{- \be f_\pm (\mu) N} }{
  \nu_- e^{- \be f_- (\mu) N} + \nu_+ e^{- \be f_+ (\mu) N} }.
\end{equation}
Namely, if neither $\la_+$ nor $\la_-$ is negligible, both phases are dominant, and we speak of a two-phase regime. On the other hand, if one of the weights is negligible, only one phase is dominant, and we speak of a single-phase regime. In transition between them yet another regime arises; we call it a crossover regime. These three regimes may be identified rather generally and not only for model~\eqref{eq: H}. This follows from the fact that Eq.~\eqref{eq: Z} for the partition function $Z$ and, hence, also Eq.~\eqref{eq: f} for the free energy $f$ are applicable for a large group of lattice-gas models (see the Introduction).

We shall consider the three regimes separately as follows. For the two-phase and crossover regimes we will first derive formulas for the dependence $D_c (\te)$ of the diffusion coefficient on the coverage, starting from Eq.~\eqref{eq: f} for $f$. Since this equation is general and since the explicit expressions~\eqref{eq: f_al} for the single-phase free energies $f_\pm$ of model~\eqref{eq: H} will not be used in the derivation, the so obtained formulas for $D_c (\te)$ are of quite universal nature. We will then apply the formulas to model~\eqref{eq: H} and compare the results with the numerical data yielded from the elimination of $\mu$ between $\te(\mu)$ and $D_c (\mu)$. In a single-phase regime, however, the free energy $f$ is practically identical to the corresponding single-phase free energy $f_\pm$. Therefore, $D_c (\te)$ in this regime can be determined only from the explicit expressions~\eqref{eq: f_al} for $f_\pm$, yielding necessarily a result that is model dependent.

\emph{Remark.} Note that weights~\eqref{eq: la} satisfy $0 < \la_\pm < 1$ and $\la_- + \la_+ = 1$. In addition, since the stable phase has the lowest specific free energy, for $\mu$ below (above) $\mu_t$ the difference $\De f$ is positive (negative), and the weight $\la_-$ ($\la_+$) approaches $1$ exponentially fast. At $\mu_t$ the specific free energies are identical, yielding $\la_\pm = \nu_\pm / (\nu_- + \nu_+)$.

\subsection{Two-phase regime}

The two-phase regime occurs when $\te (\mu)$, $\xi (\mu)$, and $\chi (\mu)$ do not reduce to their single-phase values, which is true if both $\la_- (\mu)$ and $\la_+ (\mu)$ are of order larger than $N^{-1}$ \cite{MT11}. This can hold only near the transition---for
\begin{equation} \label{eq: interval 2ph}
  \be |\mu - \mu_t| \leq \de
\end{equation}
with a small $\de > 0$. Indeed, rewrite weights~\eqref{eq: la} as $\la_\pm (\mu) = 1/\{ 1 + \rho^{\pm 1} \exp[\pm \be \De f (\mu) N] \}$, where $\rho = \nu_- / \nu_+$. Taking the Taylor expansion $\De f (\mu) = - \De\te (\mu - \mu_t) [ 1 + O(\be|\mu - \mu_t|) ]$ of $\De f$ around $\mu_t$, we see that the order of $\la_\pm$ is at least $N^{-1}$ if $\exp(\De\te N \de)$ is small compared to $N$, i.e., if
\begin{equation} \label{eq: de}
  \de = \frac\ga{\De\te} \frac{\ln N}N
\end{equation}
with a constant $1/2 < \ga < 1$. To be specific, we set $\ga = 3/4$; then the order of $\la_\pm$ is $N^{-3/4}$ or larger.

Combining Eqs.~\eqref{eq: f} and \eqref{eq: quant gen} with the Taylor expansion
\begin{equation}
  f' (\mu) = f' (\mu_t)
  + \frac{\dif f' (\mu_t)}{\dif\mu} (\mu - \mu_t)
  + O [ \be^2 (\mu - \mu_t)^2 ]
\end{equation}
(the prime denotes a derivative with respect to $\mu$ or $\ve_b$), within the two-phase interval~\eqref{eq: interval 2ph} we get
\begin{equation} \label{eq: quant 2ph}
\begin{aligned}
  \te (\mu) &= \te_- \la_- (\mu) + \te_+ \la_+ (\mu)
  + O(\de),
\\
  \xi (\mu) &= \xi_- \la_- (\mu) + \xi_+ \la_+ (\mu)
  + O(\de),
\\
  \chi (\mu) &= (\De\te)^2 \be N \la_+ (\mu) \la_- (\mu)
  [ 1 + O (N^{-1/4}) ],
\end{aligned}
\end{equation}
where $\xi_\pm = (\dif f_\pm / \dif\ve_b)_{\mu_t}$ are the single-phase average numbers (per site) of occupied bonds at the transition. Note that in the two-phase region~\eqref{eq: interval 2ph} the coverage $\te$ ranges within the interval
\begin{equation} \label{eq: interval te 2ph}
  t_- \leq \te \leq t_+,
\end{equation}
where $t_\pm = \te (\mu_t \pm \de) = \te_\pm \mp \De\te \rho^{\pm 1} N^{-3/4} + O(\de)$. Thus, in the two-phase regime the coverage attains almost all values between $\te_-$ and $\te_+$.

According to Eq.~\eqref{eq: quant 2ph}, the dependences of $\te$, $\xi$, and $\chi$ on $\mu$ in the two-phase region is primarily given by the weights $\la_\pm$. Evaluating $\la_\pm (\te)$ from the above relation for $\te(\mu)$ and the equality $\la_+ + \la_- = 1$, we readily obtain the coverage dependences of $\xi$ and $\chi$. Substituting them into Eqs.~\eqref{eq: D_c} and \eqref{eq: P}, we arrive at the result
\begin{equation} \label{eq: D_c 2ph}
  D_c (\te) \approx \frac{ D_c^0 e^{\be\mu_t} }{ \De\te N }
  \Bigl( \frac{P_-}{\te - \te_-}
  + \frac{P_+}{\te_+ - \te} + \ep \Bigr) (1 + \ep),
\end{equation}
where $P_\pm = 1 - 2\te_\pm + \xi_\pm/3$ is the probability of finding a vacant bond in phase $p_\pm$ at the transition and the error term $\ep = O(N^{3/4} \de) = O(N^{-1/4} \ln N)$. Formula~\eqref{eq: D_c 2ph} holds only for coverages in interval~\eqref{eq: interval te 2ph}.

As Eq.~\eqref{eq: D_c 2ph} shows, the coverage dependence $D_c (\te)$ of the diffusion coefficient in the two-phase regime decreases with the system size as $1/N$. For a given size, it slowly varies if $\te$ is well between $t_-$ and $t_+$, while it increases as the hyperbola $P_\pm /\De\te N |\te - \te_\pm|$ if $\te$ is close to $t_\pm$.

\begin{figure}
  \centering
  \includegraphics{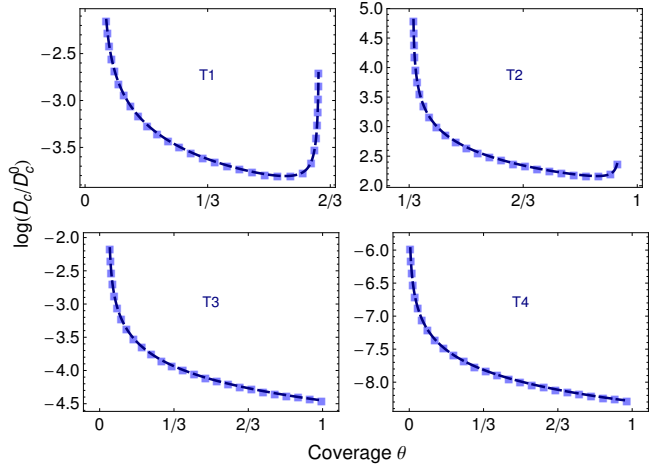}\\
\caption{The coverage dependence of the logarithm the diffusion coefficient $D_c$ (relative to $D_c^0$) in the two-phase region~\eqref{eq: interval te 2ph} for $N = 30 \times 40$. The bond interactions are $\ve_b = (-1/2) \ve_t < 0$ for T1, $\ve_b = (-4/3) \ve_t > 0$ for T2, $\ve_b = (-2/5) \ve_t > 0$ for T3, and $\ve_b = (-5/3) \ve_t < 0$ for T4. The triangle interaction $\be |\ve_t| = 4$ in all cases. The squares correspond to numerical values, whereas the dashed lines to the analytical formula~\eqref{eq: D_c 2ph}.}
  \label{fig: D_c 2ph}
\end{figure}

In order to apply the general results to model~\eqref{eq: H}, we will consider the following four representative examples of first-order phase transitions.
\begin{itemize}
  \item[(T1)]
Transition $p_0$ -- $p_{2/3}$: $\ve_t > 0$ (repulsion), $\ve_b = - \ve_t/2$ (attraction), and $|\mu - \mu_0| \leq  \ve_t$ with $\mu_0 = 3\ve_b/2$.
  \item[(T2)]
Transition $p_{1/3}$ -- $p_1$: $\ve_t < 0$ (attraction), $\ve_b = - 4\ve_t/3$ (repulsion), and $|\mu - \mu_0| \leq |\ve_t|$ with $\mu_0 = (9\ve_b + 6\ve_t)/2$.
  \item[(T3)]
Transition $p_0$ -- $p_1$: $\ve_t < 0$ (attraction), $\ve_b = - \ve_t/3$ (repulsion), and $|\mu - \mu_0| \leq |\ve_t|$ with $\mu_0 = 3\ve_b + 2\ve_t$.
  \item[(T4)]
Transition $p_0$ -- $p_1$: $\ve_t > 0$ (repulsion), $\ve_b = - 2\ve_t$ (attraction), and $|\mu - \mu_0| \leq \ve_t$ with $\mu_0 = 3\ve_b + 2\ve_t$.
\end{itemize}
In Fig.~\ref{fig: D_c 2ph} we depict $D_c (\te)$ in the two-phase interval for these transitions. The dependence $D_c (\te)$ is obtained first numerically from Eqs.~\eqref{eq: D_c}, \eqref{eq: f}, and \eqref{eq: quant gen} and then compared to values yielded by formula~\eqref{eq: D_c 2ph} with the error term $\ep$ neglected (in fact, the logarithm of $D_c$ to base $10$ is plotted for better clarity). Obviously, the analytical formula very accurately reproduces the numerical results. If we neglect thermal effects and the error term $\ep$ in Eq.~\eqref{eq: D_c 2ph}, for model~\eqref{eq: H} we can approximately write
\begin{equation}
  D_c (\te) \approx \frac{ D_c^0 e^{\be\mu_t} }N \times
  \left\{
    \begin{array}{ll}
  3/2\te & \text{transition $p_0$ -- $p_{2/3}$}, \\
  3/2(3\te - 1) & \text{transition $p_{1/3}$ -- $p_1$}, \\
  1/\te & \text{transition $p_0$ -- $p_1$}.
    \end{array}
  \right.
\end{equation}

\subsection{Crossover regimes}

At either end the two-phase region is neighbored by a crossover region in which $\te (\mu)$, $\xi (\mu)$, and $\chi (\mu)$ rapidly reduce from two-phase to single-phase values. This corresponds to a decrease in the order of either $\la_- (\mu)$ or $\la_+ (\mu)$ from above $N^{-1}$ below it \cite{MT11}. So, using the arguments leading to Eq.~\eqref{eq: de}, we get that the two crossovers take place within the regions
\begin{equation}
  \de \leq \be |\mu - \mu_t| \leq d,
  \quad
  d = \frac c{\De\te} \frac{\ln N}N,
\end{equation}
where $c > 1$. Taking $c = 5/4$, say, the order of $\la_\pm$ reduces within the crossover regions from $N^{-3/4}$ to $N^{-5/4}$. Using the upper (lower) signs for the crossover above (below) $\mu_t$, from Eqs.~\eqref{eq: f} and \eqref{eq: quant gen} we get
\begin{equation} \label{eq: quant cross}
\begin{aligned}
  \te (\mu) &= \te_\pm \pm \chi_\pm |\mu - \mu_t|
  \mp \De\te \rho^{\pm 1} e^{ -\De\te \be N |\mu - \mu_t| }
  + \ep',
\\
  \xi (\mu) &= \xi_\pm \pm \ze_\pm |\mu - \mu_t|
  \mp \De\xi \rho^{\pm 1} e^{ -\De\te \be N |\mu - \mu_t| }
  + \ep',
\\
  \chi (\mu) &= \chi_\pm + (\De\te)^2 \be N \rho^{\pm 1}
    e^{ -\De\te \be N |\mu - \mu_t| } + \be N \ep',
\end{aligned}
\end{equation}
where $\ze_\pm = (\dif^2 f_\pm / \dif\mu \dif\ve_b)_{\mu_t}$, $\chi_\pm = (-\dif^2 f_\pm / \dif\mu^2)_{\mu_t}$ are the single-phase susceptibilities at the transition, and the error term $\ep' = O(N^{-3/2})$. Thus, in the crossovers the coverage ranges within the intervals
\begin{equation} \label{eq: interval te cross}
  \tau_- \leq \te \leq t_-,
  \quad
  t_+ \leq \te \leq \tau_+
\end{equation}
with $\tau_\pm = \te (\mu_t \pm d) = \te_\pm \pm (\chi_\pm/\be) d + O(N^{-5/4})$. Intervals~\eqref{eq: interval te cross} are very narrow and concentrated around $\te_\pm$.

The $\mu$ dependences of $\te$, $\xi$, and $\chi$ in Eq.~\eqref{eq: quant cross} are essentially given only via $|\mu - \mu_t|$. Evaluating the latter [or, more conveniently, evaluating $\exp( - \De\te \be |\mu - \mu_t| N)$] as a function of $\te$, the coverage dependences of $\xi$ and $\chi$ can be deduced. Combining them with Eqs.~\eqref{eq: D_c} and \eqref{eq: P}, we get
\begin{gather} \label{eq: D_c cross}
  D_c (\te) \approx D_c^0 e^{\be\mu_t}
  \frac{ P_\pm
  + \frac{P'_\pm}{\chi_\pm} (\te - \te_\pm)
  \pm \bigl( \frac{P'_\pm}{\chi_\pm}
    - \frac{\De P}{\De\te} \bigr) \frac{\om_\pm (\te)}{C_\pm}
  + \ep' }{
  \frac1\be \chi_\pm [ 1 + \om_\pm (\te)] (1 + N \ep')}
\\
  \intertext{with}
  \om_\pm (\te) = W ( \De\te \rho^{\pm 1}
  C_\pm e^{ \mp C_\pm (\te - \te_\pm) } ),
\end{gather}
where $P'_\pm = - 2\chi_\pm + \ze_\pm/3$ represents the rate of change of $P$ with $\mu$ in a given phase evaluated at the transition, the shorthand $C_\pm = \De\te \be N / \chi_\pm$, and $W(y)$ is the Lambert function (the inverse to $y = W \exp W$). The upper (lower) signs in formula~\eqref{eq: D_c cross} correspond to the crossover around $\te_+$ ($\te_-$).

If $\te$ is close to the two-phase region (close to $t_\pm$), then $\om_\pm \approx C_\pm |\te - \te_\pm| \gg 1$ so that $D_c (\te)$ still increases as the two-phase hyperbola $P_\pm /\De\te N |\te - \te_\pm|$. On the other hand, if $\te$ is close to a single-phase region (close to $\tau_\pm$), then $\om_\pm \approx \De\te \rho^{\pm 1} C_\pm \exp(- C_\pm |\te - \te_\pm| ) \ll 1$ so that the diffusion coefficient behaves as $[ P_\pm + (P'_\pm / \chi_\pm) (\te - \te_\pm) ]/(\chi_\pm / \be)$, i.e., as a linear perturbation from the constant value $P_\pm / (\chi_\pm / \be)$. Thus, within a crossover region the diffusion coefficient suddenly changes from the hyperbolic increase to a slight linear increase (or decrease) as $\te$ moves from two-phase region across $\te_\pm$ towards a single-phase region.

\begin{figure}
  \centering
  \includegraphics{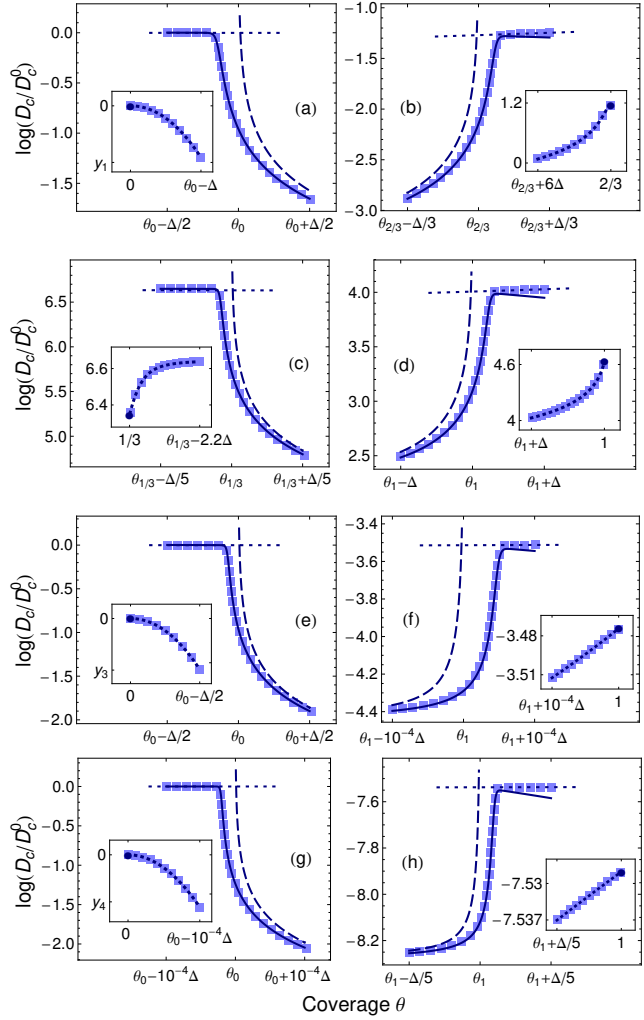}\\
\caption{The coverage dependence of the logarithm of the diffusion coefficient (relative to $D_c^0$) in the crossover and  single-phase regions for the same system size and interparticle interactions as in Fig.~\ref{fig: D_c 2ph}. The squares represent numerical values, whereas the dashed and solid lines correspond to the analytical formula~\eqref{eq: D_c 2ph} and \eqref{eq: D_c cross}, respectively. The dotted lines represent the analytical dependence obtained in the single-phase regions, and the disks depict their limiting values. The shorthands $\De = N^{-3/4}$, $y_1 = -1 \times 10^{-3}$, $y_3 = - 7 \times 10^{-4}$, and $y_4 = 1.2 \times 10^{-11}$.}
  \label{fig: D_c cross}
\end{figure}

For model~\eqref{eq: H} the dependence $D_c (\te)$ in the crossover regions for transitions~T1 -- T4 is shown in Fig.~\ref{fig: D_c cross}.

\subsection{Single-phase regimes}

Finally, far from the transition there is a single dominant phase: $p_-$ for $\mu$ below $\mu_t - d$ (i.e., $\te$ below $\tau_-$) and $p_+$ for $\mu$ above $\mu_t + d$ (i.e., $\te$ above $\tau_+$). Then $\te (\mu)$, $\xi (\mu)$, and $\chi (\mu)$ reduce to their single-phase values. Indeed, denoting the stable phase by $p_\al$, Eqs.~\eqref{eq: f} and \eqref{eq: quant gen} yield
\begin{equation} \label{eq: quant single}
\begin{aligned}
  \te (\mu) &= - \frac{\dif f_\al (\mu)}{\dif\mu} + O(N^{-5/4}),
\\
  \xi (\mu) &= \frac{\dif f_\al (\mu)}{\dif\ve_b} + O(N^{-5/4}),
\\
  \chi (\mu) &= - \frac{\dif^2 f_\al (\mu)}{\dif\mu^2}
  + O(\be N^{-5/4}).
\end{aligned}
\end{equation}

The coverage dependence of $D_c$ in this regime follows from the inverse to $\te (\mu)$ that can be obtained only from the explicit expressions~\eqref{eq: f_al} for $f_\al$. The dependence $D_c (\te)$ is simple to derive for $\al = 0,1$, while for $\al = 1/3, 2/3$ we may take into account the approximation $\ln (1+x) \approx x$ because $x \sim \exp(-\const\be)$ is small at low temperatures. In this way we arrive at the formulas
\begin{subequations} \label{eq: D_c single}
\begin{equation}
  \frac{D_c}{D_c^0} \approx
  \left\{
    \begin{array}{ll}
    (1-2\te)/(1-\te)^2 & \text{regime of $p_0$},
    \\
    (5 - 9\te + a)(3\te - 1 + a)/8 q_b^3 a
      & \text{regime of $p_{1/3}$},
    \\
    2 / q_b^3 \sqrt{ 8q_b^3 q_t^6 + (2-3\te)^2 }
      & \text{regime of $p_{2/3}$},
    \end{array}
  \right.
\end{equation}
and $D_c \approx 0$ in the regime of phase $p_1$, where $a = [8q_b^3 + (1-3\te)^2]^{1/2}$, $q_b = \exp(-\be\ve_b)$ and $q_t = \exp(-\be\ve_t)$. For phase $p_1$ the approximation of $f_1$ that uses only one-site perturbations is not sufficient to get a non-vanishing $D_c (\te)$. In order to resolve this drawback, we need to take into account the next dominant contributions arising from two-site perturbations (the removal of two particles in a bond). Then an additional term $(-3/\be) \ln[ 1 + \exp(-\be \De H_1^{\bullet\bullet}) ]$ appears in $f_1$, where $\De H_1^{\bullet\bullet} = 2\mu - 11\ve_b - 10\ve_t$ is the energy excess of a two-site perturbation of $\si_1$ over $\si_1$ \cite{MT12}. Applying this refined expression for $f_1$, we get
\begin{equation}
  D_c (\te) \approx \frac{D_c^0}
  { q_b^5 q_t^4 \sqrt{1 + 24 q_b q_t^2 (1-\te) } }
  \quad
  \text{regime of $p_1$}.
\end{equation}
\end{subequations}

The dependence $D_c (\te)$ in the single-phase regions for transitions~T1 -- T4 is detailed in the insets in Fig.~\ref{fig: D_c cross}. It turns out that $D_c(\te)$ does not diverge in the single-phase regimes as might be incorrectly conjectured from the behavior of $D_c$ in the two-phase and crossover regions. Rather, it tends to the constant values $1$, $(1 + \sqrt{2 q_b^3})/4 q_b^3$, $1/\sqrt{2 q_b^9} q_t^3$, and $1/q_b^5 q_t^4$ as $\te$ approaches $0$, $1/3$, $2/3$, and $1$, respectively. However, the model does not allow us to analyze the behavior of $D_c (\te)$ on both sides of $\te = 1/3, 2/3$ due to the infinite number of ground states on the lines separating the regions of $\si_0, \si_{1/3}$ and $\si_{2/3}, \si_1$.

\section{Conclusions and final remarks}

We have investigated the dependence $D_c (\te)$ of the chemical diffusion coefficient on the surface coverage at low temperatures, assuming that a first-order phase transition between two phases takes place in the system and that the local limit approximation is applicable. Our analysis was based on an expression for $D_c$, Eq.~\eqref{eq: D_c}, available within this approximation and on a general formula for the finite-size specific free energy $f$, Eq.~\eqref{eq: f}, valid near such a transition. The key aspect of the approximation was that $D_c$ could be evaluated only from $f$. Hence, rather crudely but plausibly, the original kinetic problem was reduced to a thermodynamic one.

We identified three types of regions each of which was associated with a different behavior of $D_c (\te)$: a two-phase region at or very close to the transition, two single-phase regions farther away from the transition, and two crossover regions in between. Combining the expression for $D_c$ and the formula for $f$, we derived rather universal finite-size formulas for $D_c (\te)$ in the two-phase and crossover regions, Eqs.~\eqref{eq: D_c 2ph} and \eqref{eq: D_c cross}, and applied them to an illustrative model of surface diffusion on a triangular lattice (see Figs.~\ref{fig: D_c 2ph} and \ref{fig: D_c cross}). However, in a single-phase region it was possible to obtain only model-dependent formulas.

It should be true that the analytical formula for $D_c (\te)$ valid in one of the three regions quite smoothly takes over from the formula valid in the neighboring region. Figure~\ref{fig: D_c cross} shows that this requirement is clearly satisfied. Moreover, as may be expected, the agreement between different formulas increases quite fast as the system size $N$ grows.

The correlation factor $P$ that appears in the approximate expression for $D_c$, Eq.~\eqref{eq: D_c}, is connected with the interactions of an activated adparticle with its neighbors. We considered the simplest situation, Eq.~\eqref{eq: P}, when these interactions were neglected so that $P$ was the probability of finding a vacant bond. Nevertheless, our main results---formulas~\eqref{eq: D_c 2ph} and \eqref{eq: D_c cross} for $D_c (\te)$---remain valid in the two-phase and single-phase regions also for the general version of $P$ given by Eq.~\eqref{eq: P gen}. Indeed, to obtain these formulas, we only used that $P$ was a sum of first derivatives of the free energy $f$. In the simplest form of $P$ these derivatives were $\te$ and $\xi$, the latter one being an example of such a derivative in general. It is obvious from Eqs.~\eqref{eq: quant 2ph} and \eqref{eq: quant cross} that the $\mu$ dependences of $\te$, $\xi$, or any other first derivative $\dif f/\dif p_i$ of $f$ in each of the three regimes have exactly the same form. Therefore, since a general $P$ is a sum of such derivatives, its coverage dependence must also have the same form as in the simplest version considered in this paper. As a result, our formulas for $D_c (\te)$ can be readily extended to the general $P$ given by Eq.~\eqref{eq: P gen} simply by setting $P_\pm = a_0 + \sum_i a_i (\dif f_\pm/\dif p_i)_{\mu_t}$ and $P'_\pm = \sum_i a_i (\dif^2 f_\pm/\dif\mu \dif p_i)_{\mu_t}$ in these formulas. Again, $P_\pm$ are the single-phase values of $P$ and $P'_\pm$ are the single-phase rates of changes of $P$ with $\mu$, both evaluated at the transition.

It should be realized, however, that the extension to a general $P$ is possible only if the partition function of a given model can be written as in Eq.~\eqref{eq: Z}. Even though the equation is true for a large class of models, its applicability to a particular model must be always verified, similarly to the low-temperature analysis described in Sec.~\ref{sec: model}. Moreover, it may be necessary to include additional terms in the model Hamiltonian so that all of the parameters $p_i$ are present and the derivatives $\dif f/\dif p_i$ and $\dif^2 f_\pm/\dif\mu \dif p_i$ can be obtained (perhaps setting $p_i$ equal to zero in the end).

Finally, note that the formula for $D_c (\te)$ in the two-phase regime, Eq.~\eqref{eq: D_c 2ph}, can approximately yield the Langmuir relation, $D_c (\te) = \const/(1-\te)$, only if phase~$p_+$ is fully occupied, its factor $P_+$ has a strictly positive value, and the factor $P_-$ in the other phase is negligible. This is not the case for the simplest version of $P$ given by Eq.~\eqref{eq: P}.


\section*{Acknowledgments}

This research was supported by the Czech Science Foundation, Project No.~P105/12/G059.


\end{document}